\documentclass[a4paper,11pt]{article}
\usepackage{amsfonts}
\usepackage{latexsym}
\usepackage{amsmath}
\usepackage{amssymb}
\usepackage{amssymb}
\usepackage{slashed}
\usepackage{upgreek}
\usepackage{mathrsfs}
\usepackage{bbm}

\usepackage{amsthm}

\theoremstyle{remark}

\usepackage[utf8]{inputenc}
\usepackage[english]{babel}

\theoremstyle{definition}

\allowdisplaybreaks

\usepackage{relsize}
\usepackage{graphicx}
\usepackage{comment}

\renewcommand{\thefootnote}{\fnsymbol{footnote}}

\def\appendix#1{\addtocounter{section}{1}\setcounter{equation}{0}
\renewcommand{\thesection}{\Alph{section}}
\section*{Appendix \thesection\protect\indent \parbox[t]{11.15cm}{#1}}
\addcontentsline{toc}{section}{Appendix \thesection\ \ \ #1}}

\font\mybb=msbm10 at 11pt

\def\bb#1{\hbox{\mybb#1}}

\def\bZ {\bb{Z}}

\def\bR {\bb{R}}

\def\h{\widehat}
\def\be{\begin{equation}}
\def\ee{\end{equation}}

\def\rsq {]\kern -2.0pt]}
\def\lsq {[\kern -2.0pt[}

\newcommand{\bea}{\begin{eqnarray}}
\newcommand{\eea}{\end{eqnarray}}

\usepackage[usenames,dvipsnames,svgnames,table]{xcolor}	
\usepackage[normalem]{ulem}				
\usepackage{microtype}
\usepackage[colorlinks, citecolor=blue, linkcolor=blue, urlcolor=Maroon, filecolor=Maroon, linktocpage=true]{hyperref} 

\usepackage{chngcntr}
\counterwithout{equation}{section}

\begin{document}

\begin{center}
\vspace*{-1.0cm}
\begin{flushright}
\end{flushright}


\vspace{2.0cm} {\Large \bf Heterotic horizons and  AdS$_3$ backgrounds that preserve 6 supersymmetries} \\[.2cm]

\vskip 2cm
 Georgios  Papadopoulos
\\
\vskip .6cm


\begin{small}
\textit{Department of Mathematics
\\
King's College London
\\
Strand
\\
 London WC2R 2LS, UK}\\
\texttt{george.papadopoulos@kcl.ac.uk}
\end{small}
\\*[.6cm]

\end{center}

\vskip 2.5 cm

\begin{abstract}
\noindent

We prove, under suitable global assumptions, that the only heterotic horizons with closed 3-form field strength that preserve strictly 6 supersymmetries have spatial horizon section diffeomorphic to either $SU(3)$ or $S^2\times S^3\times SO(3)$, up to identifications with the action of a discrete group. Under similar assumptions,  which include the compactness of the transverse space, we demonstrate that there are no heterotic AdS$_3$ solutions that preserve  6 supersymmetries. The proof is based on a topological argument.

We also re-examine the conditions required for the existence of such backgrounds that preserve 4 supersymmetries focusing on those that admit an additional $\oplus^2\mathfrak{u}(1)$ symmetry. We provide some additional explanation for the existence of solutions and   point out the similarities that these conditions have with those that have recently emerged in  the classification of compact strong 6-dimensional Calabi-Yau manifolds with torsion.

\end{abstract}



\newpage


\numberwithin{equation}{section}

\renewcommand{\thefootnote}{\arabic{footnote}}



\section{Introduction}\label{intro}

The last few years substantial progress has been made to identify  the geometry of supersymmetric backgrounds of supergravity theories, for a review see e.g. \cite{ugjggp} and references therein. This is especially the case for those that have applications in supergravity and string compactifications, see e.g. \cite{FR, romans1, romans2, Pope:1984} and reviews \cite{duff, grana};  in AdS/CFT correspondence \cite{maldacena} and related work in e.g. \cite{gibbons}-\cite{nmar};  and in black holes, see e.g. \cite{israel}-\cite{robinson} for early works on uniqueness in four dimensions  and in e.g. \cite{gibbonstownsend}-\cite{kl} for later developments in higher dimensions.  Such an endeavour involves the solution of the Killing spinor equations (KSEs) of the associated supergravity theories, which in turn describe the geometry of supersymmetric backgrounds. The KSEs imply
some of the field equations of these theories but not all. The remaining field equations and Bianchi identities of the form field strengths  are a set of typically non-linear partial differential equations (PDEs) that have to be imposed in addition to the KSEs. Apart from some special cases, usually involving solutions that preserve a large number of supersymmetries, see e.g. \cite{gpjfof}, the solution of these remaining PDEs  is challenging and the main obstacle to proceed from the identification of geometry of supersymmetric backgrounds to their eventual classification.  Nevertheless, many special solutions are known based on the use of ansatze to simply the equations.   Such additional assumptions are useful for  applications to physics but they do not give the most general solution to the problem.

Focusing on the heterotic\footnote{We shall take the 3-form field strength to be closed, $dH=0$ and we shall not explore the $dH\not=0$ case that arises whenever there is a non-trivial contribution from the heterotic anomaly.  So the results also apply to the common sector of type II supergravity theories.  We use the term heterotic to characterise these backgrounds  because the analysis that follows deals with only half of the KSEs of the common sector, i.e. those of the heterotic theory. }
 theory, it has been pointed out that black hole horizons with compact spatial section and a non-trivial 3-form flux can preserve 2, 4, 6 and 8 supersymmetries and those with 8 supersymmetries have been classified \cite{jggphh}.  Similarly, it is known that the heterotic theory admits  only supersymmetric AdS$_3$ backgrounds.  Furthermore, if the transverse space is compact, such backgrounds  preserve 2, 4, 6 and 8 supersymmeries and those that preserve 8 supersymmetries are locally isometric to AdS$_3\times S^3\times M^4$, where $M^4$ is a hyper-K\"ahler manifold with the radii of AdS$_3$ and $S^3$ equal \cite{sbjggp}.

In this paper, we shall extend the classification results for horizons and AdS$_3$ backgrounds that preserve 8 supersymmetries to prove a uniqueness theorem for heterotic horizons that preserve 6 supersymmetries. In particular, we shall show that if the spatial horizon section is {\it compact}, the $\mathfrak{u}(2)$ symmetry\footnote{This symmetry is generated by $\h\nabla$-covariantly constant vector fields, which do not have fixed points, and so it is nearly free up to possible identifications with the action of a discrete group, where $\h\nabla$ is the connection with torsion the 3-form field strength of the theory.} of these backgrounds has closed orbits\footnote{If the orbits of the $\mathfrak{u}(2)$ algebra are closed, the action can be integrated to an action of either a $U(2)$ or $U(1)\times SO(3)$ group -- $S(U(1)\times U(2))$ is isomorphic to $U(2)$ but the former is more convenient to use in the context of heterotic horizons.} and can be integrated to a {\it free} $S(U(1)\times U(2))$ or $U(2)$ action with {\it simply connected} space of orbits, then the only horizon that preserves strictly 6 supersymmetries has {\it spatial horizon section that is diffeomorphic to } $SU(3)$. More possibilities arise upon an identification with the action of a discrete subgroup on $SU(3)$.  The proof is based on the topological argument, which is similar to that used in \cite{gp0} for the classification of 8-dimensional strong hyper-K\"ahler manifolds with torsion (HKT). Though, the  geometric properties of the spaces used in the proof have a  different origin in the two cases.

In addition, we shall  consider the case that the $\mathfrak{u}(2)$ symmetry of heterotic horizons that preserve 6 supersymmetries can be integrated to a free $U(1)\times SO(3)$ action with a simply connected space of orbits. We shall demonstrate that  a topological argument similar to that employed in the $SU(3)$ case above leads to the possibility that there may be heterotic horizons diffeomorphic to $S^2\times S^3\times SO(3)$. However, unlike the $SU(3)$ case, it is not known whether $S^2\times S^3 \times SO(3)$ admits a metric and a 3-form that satisfy  all the conditions required for the existence of a heterotic horizon solution.

We shall also investigate the existence of AdS$_3$ solutions that preserve strictly 6 supersymmetries. We find  that if the transverse space is compact and the $\mathfrak{su}(2)$ symmetry acting on the transverse space can be integrated to a {\it free} $SU(2)$ or $SO(3)$ action with a {\it simply connected space of orbits}, then {\it there are no } AdS$_3$ solutions that preserve strictly 6 supersymmetries. Again, the argument  is essentially topological  -- we do not solve a PDE.

We shall also re-examine the conditions on the geometry of heterotic horizons and AdS$_3$ backgrounds that preserve 4 supersymmetries. Although all the conditions to find solutions are known \cite{jggphh}, we shall limit the discussion here to a special case that the  spatial  horizon section and the AdS transverse space admit an additional $\oplus^2\mathfrak{u}(1)$ symmetry.  This symmetry does not arise as a consequence of the KSEs. Nevertheless,  such backgrounds have been examined before in \cite{jggp2} to investigate the conditions for the existence of solutions. We clarify several aspects  of the construction and point out that the differential system that emerges is closely related to that recently derived in \cite{abls} for the classification of 6-dimensional compact strong  Calabi-Yau manifolds with torsion. After exploring the associated differential system, we find that to classify such backgrounds that preserve 4 supersymmetries, in addition to a {\it topological condition}, one has to solve a {\it non-linear PDE}.  This PDE takes the form
\be
\mathring{\nabla}^2 u= e\, u^2-p(x)~,
\label{pde}
\ee
where $e>0$ is a constant and $p(x)>0$ is a positive function that may depend on the Ricci tensor of $M^4$. In the applications at hand, $u$ is the  scalar curvature of a 4-dimensional K\"ahler manifold $M^4$ with metric $\mathring g$,  $u=R(\mathring g)$ and $\mathring{\nabla}$ is the Levi-Civita connection.  The same PDE occurs in the classification of heterotic horizons that preserve 6 supersymmetries and  8-dimensional compact strong HKT manifolds \cite{gp0}.  In this case $u$ is the scalar curvature $R$ of a anti-self-dual 4-dimensional manifold $M^4$.  Thus, this PDE occurs in a variety of problems. Although examples of solutions to this PDE are known, to our knowledge, {\it there is not a general theory} for establishing either existence or uniqueness of solutions\footnote{It has been pointed out in \cite{gp0} that it can always be solved for $u$ given a fixed metric on $M^4$.} for the metric $\mathring g$ on $M^4$.  We shall also point out that to find all solutions, one  has to also consider appropriate covers of the solutions obtained after solving all the conditions that arise from the field equations, KSEs and Bianchi identities. This point is illustrated with an example.

This paper is organised as follows. In section two, we begin with a summary  of the geometry of supersymmetric heterotic horizons and especially those   that preserve 6 supersymmetries. Then,  we provide the argument to prove that if the horizon sections are compact and can be considered as principal bundles with fibre
$S(U(1)\times U(2))$ or $U(2)$ over a simply connected 4-dimensional manifold $M^4$, then the only horizons that preserve 6 supersymmetries are those with spatial horizon sections diffeomorphic to $SU(3)$, where $M^4=\overline{\mathrm{CP}}^2$.  This is the complex projective space with orientation opposite to that given by the standard complex structure. The proof is topological in nature and utilises the cohomological consequences that the closure of the 3-form field strength has on $M^4$. In particular, the closure of the 3-form field strength is expressed as a condition that involves the Euler and signature classes of $M^4$ and
 the first Chern class of a line bundle, see (\ref{top3}) below.

 Moreover, we shall also consider the case that the spatial horizon section is a principal bundle with fibre group $U(1)\times SO(3)$ with a simply connected base space $M^4$. We shall show, using a similar topological argument, that the spatial horizon section is diffeomorphic to $S^2\times S^3\times SO(3)$. It turns out that $M^4=\#_2 \overline{\mathrm{CP}}^2$. However, unlike the $SU(3)$ case, it is not known whether   $S^2\times S^3\times SO(3)$ admits any metric that solves all the conditions required for this space to be a heterotic horizon solution.

 In section three, after a summary of the geometry of supersymmetric heterotic AdS$_3$ backgrounds,  we present a similar analysis for the existence of  smooth solutions  with compact transverse space $M^7$ that is a principal bundle over a simply connected base space $M^4$. Again, the closure of the 3-form field strength imposes a cohomological condition given in (\ref{top4}) below. Unlike, the horizons' case, the restrictions on the geometry of $M^4$ imposed by the KSEs lead to the conclusion that the cohomological condition has no solutions.  Thus, there are no such smooth AdS$_3$ backgrounds that preserve strictly 6 supersymmetries.

 In section 4, we revisit the conditions for the existence of heterotic horizons and AdS$_3$ backgrounds that preserve 4 supersymmetries. We explore again the coholomogical consequences of the closure of the 3-form field strength and explain the role that (\ref{pde}) has for such solutions to exist.

\section{Heterotic horizons}

\subsection{Summary of the geometry of supersymmetric heterotic horizons}

The details of the description of geometry of heterotic horizons, which includes a proof of the key statements,  can be found in \cite{jggphh}. Here,
we shall just state those results that we shall use later. The metric $g$ and 3-form field strength $H$ of the 10-dimensional spacetime are given by
\be
g= 2 e^+ e^-+ \tilde g~,~~~H=d (e^-\wedge e^+)+ \tilde H~,~~~d\tilde H=0~,
\label{horgh}
\ee
where
\be
e^+=du~,~~~e^-=dr+r h~,
\label{horgh1}
\ee
where $u,r$ are  spacetime (light-cone) coordinates,  $\tilde g$ and $\tilde H$ is the metric and 3-form on the 8-dimensional  spatial horizon section $\mathcal {S}$. In particular, they depend only on the coordinates of $\mathcal {S}$. Moreover $h$ is an 1-form on $\mathcal {S}$ and,  similarly, the dilaton $\Phi$ is a function on $\mathcal {S}$.

An investigation of the KSEs and the use that $\mathcal{S}$ is taken to be compact space for (black hole) horizons reveals that $h$ is $\h{\tilde \nabla}$ covariantly constant and the associated vector field leaves $\Phi$ invariant, i.e.
\be
\h{\tilde \nabla} h=0~,~~~ h\Phi=0~,
 \ee
 where $\h{\tilde \nabla}$ is the metric connection on $\mathcal{S}$ with torsion $\tilde H$ and we denote with the same symbol the 1-form $h$ and the associated vector field,  $h(X)=\tilde g(h, X)$ or equivalently $h=\tilde g^{-1} h$. Schematically, $\h{\tilde \nabla}=\tilde \nabla+\frac{1}{2} \tilde g^{-1} \tilde H$,  where $\tilde \nabla$ is the Levi-Civita connection of $\tilde g$.
 As a result, the first condition above implies that $h$ is Killing vector field on $\mathcal{S}$ and $dh=\iota_h \tilde H$, where $\iota$ denotes inner derivation operation with respect to a vector field. This in turn implies that $\tilde H$ is also invariant as $d\tilde H=0$, $\mathcal{L}_h \tilde H=0$.

 Furthermore, it can be shown that heterotic horizons preserve 2, 4, 6 and 8 supersymmetries. In fact for 2 supersymmetries, the holonomy of $\h{\tilde \nabla}$ is contained in $G_2$.  The geometry of $\mathcal{S}$ is further refined, if the horizon preserves additional supersymmetries, like 4,6 and 8. These conditions will be summarised later for 4 and 6 supersymmetries.  In what follows, we shall need the expression for the scalar curvature of $\mathcal{S}$. This can be  written as
\be
R(\tilde g) = \frac{1}{4}  \tilde H^2 -2  \tilde \nabla^2\Phi
~,
\label{efe8}
\ee
which follows after a substitution of  (\ref{horgh} into the Einstein equations of the heterotic theory , see \cite{jggphh} for a derivation.

\subsection{Heterotic horizons with 6 supersymmetries}

\subsubsection{Summary of the geometry}

For horizons that preserve 6 supersymmetries, it has been demonstrated in \cite{jggphh} that the spatial horizon section $\mathcal{S}$  is an 8-dimensional strong\footnote{This means that the torsion $\tilde H$ of the manifold is a closed 3-form, $d\tilde H=0$.} HKT manifold  that admits a symmetry with Lie algebra
$\mathfrak{u}(2)=\mathfrak{u}(1)\oplus \mathfrak{su}(2)$ -- the description of the HKT geometry has originally be given in \cite{hkt} and for a more recent review see \cite{gpew}. This means  the spatial horizon section $\mathcal {S}$ (and the spacetime) admit four (linearly independent) vector fields\footnote{Compact HKT manifolds that are not conformally balanced admit a $\mathfrak{u}(2)$ symmetry generated by $\h\nabla$-covariantly constant vector fields. The proof of the existence of such an action is rather involved and utilises the work of Perelman as applied to generalised Ricci flows.
However here, the existence of such vector fields is a consequence of the KSEs and of a global argument for which the field equations of the theory are used, see \cite{jggphh}.} that are Killing and leave $\tilde H$ and $\Phi$ invariant. In addition, the associated 1-forms $\{h^0, h^r; r=1,2,3\}$  on $\mathcal{S}$ are $\h{\tilde \nabla}$-covariantly constant 1-forms   with $h^0=h$ as in (\ref{horgh1}). The vector field associated  to $h^0$ is generated by the action of the $\mathfrak{u}(1)$ subalgebra and so it commutes with the rest of the vector fields. The length of these 1-forms is constant and we take them to be orthogonal of length $k$, i.e. $h^2=k^2$ and similarly for the rest. Next define $\{\lambda^a ; a=0,1,2,3\}$ with $\lambda^0= k^{-1} h$ and $\lambda^r= k^{-1} h^r$ to emphasize that these later will be declared to be components of a principal bundle connection with gauge Lie algebra $\mathfrak{u}(1)\oplus \mathfrak{su}(2)$. Then, the metric and torsion of $\mathcal {S}$
can be written as
\be
\tilde g=\delta_{ab} \lambda^a \lambda^b+ e^{2\Phi} \mathring g~,~~~\tilde H=\mathrm{CS}(\lambda)-\mathring{*}^4 d e^{2\Phi}~,
\label{gh6}
\ee
where $\mathring g$ is a metric on the space of orbits of the $\mathfrak{u}(1)\oplus \mathfrak{su}(2)$ action,  $M^4$ and $\mathrm{CS}(\lambda)$ stands for the Chern-Simons 3-form of $\lambda$.

 From now on, let us assume\footnote{It is known that if the vector fields generated by the action of a Lie algebra on a manifold are complete, then, according to Palais theorem, this action can be integrated to a group action, where the group is the unique simply connected group with that Lie algebra. In the case at hand, the $\mathfrak{u}(2)$ action on $\mathcal{S}$ can be integrated to an action of $\bR\times SU(2)$. As the vector fields are no-where zero, because they are covariantly constant, this is an almost free action. However, the orbits of $\bR$ subgroup may not be closed and potentially dense in the space. To avoid this, we assume that they are closed and so the whole action can be integrated to an action of either $U(2)$ or $S(U(1)\times U(2))$. Still, this action may not be free -- there may be identifications with a discrete subgroup and the space of orbits can be an orbifold. So we shall assume that $\mathcal {S}$ is a principal bundle as stated up to an identification with a discrete group. For a more general set up than principal bundles, one can consider the theory of foliations, see \cite{bfgv} for a careful explanation of these issues involved in the context of 8-dimensional HKT manifolds.} that $\mathcal{S}$ is either a $S(U(1)\times U(2))$ or a $U(2)$ principal bundle over  $M^4$ with connection $\lambda$. Then, it can be shown, see \cite{jggphh},  that the base space of the fibration $(M^4, \mathring {g}, \mathring{I}_r)$ is a quaternionic K\"ahler manifold with quaternionic structure $\{\mathring{I}_r; r=1,2,3\}$. In particular,
\be
\mathring{\nabla} \mathring{I}_r+\frac{k}{2} \xi^t \epsilon_t{}^s{}_r \mathring{I}_s=0~,
\label{qk2}
\ee
where $\mathring \nabla$ is the Levi-Civita connection of $\mathring g$,  $\xi^r$ is the pull-back of the connection $\lambda^r$ on $M^4$ with a local section and $r,s,t=1,2,3$.
As we shall see below the geometry of $M^4$ is restricted further.  The curvature, ${\mathcal F}$, of $\lambda$ satisfies the conditions
\be
\mathcal {F}^0= (\mathcal {F}^0)^{\mathrm{asd}}~,~~~(\mathcal {F}^r)^{\mathrm{sd}}= \frac{k}{4} e^{2\Phi} \mathring \omega^r~,
\label{concomp}
\ee
 where $(\mathcal {F}^a)^{\mathrm{asd}}$ denotes the anti-self-dual\footnote{The orientation chosen is that of the quaternionic K\"ahler structure  on $M^4$, i.e. the anti-self-dual component of a 2-form $\chi$ is identified with the $(1,1)$ and $\mathring\omega^r$-traceless component of $\chi$ with respect to any of $\mathring I_r$.} component  of $\mathcal{F}^a$ while the self-dual component of $(\mathcal {F}^r)^{\mathrm{sd}}$ is restricted as indicated.  $\mathring \omega^r$ are the Hermitian forms of the quaternionic K\"ahler structure with respect to the metric $\mathring g$.   This summarises the geometry of $\mathcal{S}$ for horizons that preserve 6 supersymmetries.

\subsection{Uniqueness of horizons with 6 supersymmetries}\label{sec:hor}

The line of argument that follows to prove the uniqueness of heterotic horizons that preserve 6 supersymmetries is related to that given in \cite{gp0} for the uniqueness of 8-dimensional compact strong HKT manifolds. For completeness, the main steps with some alterations are summarised below.
It is clear from (\ref{gh6}) that the closure of $\tilde H$, $d \tilde H=0$,  requires for consistency that the cohomology class
$[\mathcal{F}\wedge \mathcal{F}]$ of the 4-form $\mathcal{F}\wedge \mathcal{F}$ must be trivial, i.e.
\be
[\mathcal{F}\wedge \mathcal{F}]\equiv [\delta_{ab} \mathcal{F}^a\wedge\mathcal{F}^b]=0~,
\label{top2}
\ee
where $\mathcal{F}$ is the curvature of $\lambda$.
In addition, the differential condition $d \tilde H=0$ can be rewritten as
\begin{align}
\mathring{\nabla}^2 e^{2\Phi}&= \frac{3 k^2 }{8}  e^{4\Phi}-\frac{1}{2} \Big((\mathcal{F}^0)_o^2+ \sum_r \big((\mathcal{F}^{\mathrm {asd}})^r\big)_o^2\Big)~,
\label{dconx8}
\end{align}
where, as indicated, the inner products in the right hand side have been taken with respect to the metric $\mathring g$.  Clearly, as a differential equation on $e^{2\Phi}$, it is of the form (\ref{pde}) with $u=e^{2\Phi}$.

After some computation that has been explained in more detail in \cite{gp0}, a comparison of (\ref{efe8}) with (\ref{dconx8}) reveals that
\be
\mathring R\equiv R(\mathring g)=\frac{3k^2}{2} e^{2\Phi}>0~.
\label{posR}
\ee
As a result of (\ref{concomp}) and (\ref{posR}), see proposition 7.1 page 92 in \cite{salamon},  $M^4$ is an anti-self-dual 4-manifold with positive Ricci scalar. Since we have assumed that $M^4$ is simply connected as well,  it is a consequence of the results of \cite{lebrun0, lebrun} that $M^4$ is homeomorphic to either the connected sum $\#_n \overline{\mathbb{C}\mathrm{P}}^2$ or $S^4$, i.e.
\be
M^4= \#_n \overline{\mathbb{C}\mathrm{P}}^2~~\mathrm{or}~~S^4~,
\label{consum}
\ee
where $\overline{\mathbb{C}\mathrm{P}}^2$ denotes $\mathbb{C}\mathrm{P}^2$ equipped with the opposite orientation to that given by the standard complex structure.

Before, we proceed further, we shall point out that (\ref{dconx8}) can be written in terms of the Ricci tensor of $\mathring g$. For this, consider the integrability condition of (\ref{qk2}) to find
\be
\mathring R_{ij}{}^k{}_m \mathring I_r{}^m{}_p -\mathring R_{ij}{}^m{}_p \mathring I{}_r{}^k{}_m+\frac{k}{2} \mathcal{F}_{ij}^t\epsilon_t{}^s{}_r \mathring I_s{}^k{}_p=0~,
\ee
where $i,j,k, m, p=1,\dots, 4$,  $r=1,2,3$ and $\mathring R$ is the Riemann curvature of $\mathring g$.
Acting on this expression with $\mathring I{}_{r'}{}^p{}_q$ and after using the algebraic properties of the quaternionic structure, we deduce that
\be
\mathring R_{ij}{}^k{}_m (-\delta_{rr'} \delta^m{}_q+ \epsilon_{rr'}{}^{s'} \mathring I_{s'}{}^m{}_q)+ \mathring R_{ij mp} \mathring I{}_r{}^{mk} \mathring I_{r'}{}^p{}_q+\frac{k}{2} \mathcal{F}^t_{ij}\, \epsilon_t{}^s{}_r (-\delta_{sr'} \delta^k{}_q+\epsilon_{sr'}{}^{s'} \mathring I_{s'}{}^k{}_p)=0~.
\ee
Summing over $r$ and $r'$, this equation becomes
\be
-3 \mathring R_{ij}{}^k{}_q+\sum_r \mathring R_{ijmp}\mathring I{}_r{}^{mk} \mathring I_{r}{}^p{}_q+k \mathcal{F}^s_{ij} \mathring I_s{}^k{}_q=0~.
\ee
Using the identity
\be
\sum_r \mathring I_r{}^{mk} \mathring I_r{}^{pq}= \mathring \epsilon^{mkpq}+ \mathring g^{mp} \mathring g^{kq}-\mathring g^{kp} \mathring g^{mq}~,
\ee
and acting with $\delta^i_k$, we find,  after a re-arrangement of indices, that
\be
\mathring R_{ij}-\frac{k}{2} \mathcal {F}^r_{ki} \mathring I_r{}^k{}_j=0~,
\ee
where for simplicity  we have set $\mathring R_{ij}\equiv R(\mathring g)_{ij}$.
In turn, this and (\ref{concomp}) imply that
\be
\mathring R_{ij}-\frac{1}{4}\,\mathring g_{ij} \mathring R-\frac{k}{2} (\mathcal {F}^r)^\mathrm{asd}_{ki}\, \mathring I_r{}^k{}_j=0~.
\ee
Therefore, the anti-self-dual part of $\mathcal{F}^r$ is determined by the traceless part of the Ricci tensor, see also proposition 7.1 page 92 in \cite{salamon}.
As a result, the equation (\ref{dconx8}) can be rewritten as
\be
\mathring\nabla^2 \mathring R=-\frac{3 k^2}{4} (\mathcal{F}^0)_o^2- 3 \mathring R_{ij}\mathring R^{ij} +\mathring R^2~,
\label{dconx8b}
\ee
where the indices have been raised with $\mathring g$.
This equation should be viewed as an equation for the metric $\mathring g$ that it is required to have a solution for the existence of heterotic horizon solutions that preserve 6 supersymmetries. It is clearly of the form (\ref{pde}) and non-linear.

\subsubsection{Fibre group $S(U(1)\times U(2))$ or $U(2)$}

Suppose that $\mathcal{S}$ is a principal bundle over $M^4$ with fibre group either $S(U(1)\times U(2))$ or $U(2)$. Then, it follows from (\ref{qk2}) that the associated vector bundle of $\mathcal{S}$ with respect to the adjoint representation must be identified with the bundle of self-dual 2-forms on $M^4$, $\Lambda^+(M^4)$, i.e.
\be
\mathrm{Ad}(\mathcal{S})=\Lambda^+(M^4)~.
\ee
 Principal bundles over $M^4$ with fibre group  $S(U(1)\times U(2))$ -- the analysis of those with fibre group $U(2)$ is similar -- are associated with a complex vector bundle $E$ and a complex line bundle $L$ whose first Chern classes,  $c_1(E)$ and $c_1(L)$, respectively, satisfy the relation
\be
c_1(E)+c_1(L)=0~.
\label{trcon}
\ee
  This relation is required because  the fibre group of the principal bundle $\mathcal {S}$ is special. As a result, such principal bundles are classified by  $(c_1(L), c_2(E))\in H^2(B^4, \bZ)\oplus H^4(B^4, \bZ)$, where $c_2(E)$ is the second Chern class of $E$.

After an overall normalisation\footnote{Typically, the normalisation of the first Pontryagin class expressed in terms of principal bundle connections is $1/4\pi^2$ and it is the same as that of the square of first Chern Classes.}, the class $[\mathcal{F}\wedge \mathcal{F}]$ can be expressed as
\be
[\mathcal{F}\wedge \mathcal{F}]=c_1(L)^2+p_1(E)~,
\ee
where $p_1(E)$ is the first Pointryagin class of $E$. Using the Hirzenbruch's signature theorem
\be
p_1(\mathrm{Ad}(E))=p_1(\Lambda^+)=2\chi+3\tau~,
\label{pone}
\ee
and the classic formula in characteristic classes
\be
p_1(\mathrm{Ad}(E))=c_1(E)^2-4 c_2(E)~,
\label{classf}
\ee
which relates the first Pontryagin class of $\mathrm{Ad}(E)$ in terms of the Chern classes of $E$, we conclude that the triviality of the class $[\mathcal{F}\wedge \mathcal{F}]$ can be re-expressed as
\be
(3 c_1(L)^2+ 2\chi+3\tau)[M^4]=0~,
\label{top3}
\ee
where $\chi$ and $\tau$ are the Euler and signature characteristic classes of $M^4$, respectively,  see \cite{gp0} for more details -- the insertion of $[M^4]$ in the equation (\ref{top3}) denotes the integration of the characteristic classes over $M^4$. Note also the relation,
$p_1(E)=c_1(E)^2-2 c_2(E)$, between Pontryangin and Chern classes.  For a given $M^4$, the Euler number and signature are specified. So the only variable left to tune in order   the condition (\ref{top3}) to be satisfied  is $c_1(L)$.

It is clear that the formula (\ref{top3})  cannot be satisfied for $M^4=S^4$ because the signature of $S^4$ is zero, the Euler number is 2 and there are no non-trivial complex line bundles on $S^4$ as $H^2(S^4, \bZ)=0$.  On the other hand\footnote{The non-vanishing cohomology groups of $\#_n\overline{\mathbb{C}\mathrm{P}}^2$ are $H^0(\#_n\overline{\mathbb{C}\mathrm{P}}^2, \bZ)=H^4(\#_n\overline{\mathbb{C}\mathrm{P}}^2, \bZ)=\bZ$ and $H^2(\#_n\overline{\mathbb{C}\mathrm{P}}^2, \bZ)=\oplus^n \bZ$.}
\be
(2\chi+3\tau)[\#_n\overline{\mathbb{C}\mathrm{P}}^2]=4-n~,
\label{rhsn}
 \ee
 as the Euler number is $\chi[\#_n\overline{\mathbb{C}\mathrm{P}}^2]=2+n$ and the signature is $\tau[\#_n\overline{\mathbb{C}\mathrm{P}}^2]=-n$.  Thus, there is only one possibility that of $n=1$ with $c_1(L)^2[M^4]=-1$ -- note that $L$ admits a anti-self-dual connection. The associated principal fibration is $S(U(2)\times U(1))\hookrightarrow SU(3)\rightarrow \overline{\mathbb{C}\mathrm{P}}^2$. Thus $\mathcal {S}$ is diffeomorphic\footnote{According to Borel–Hsiang–Shaneson–Wall theorem simply connected compact finite dimensional groups admit a unique differential structure compatible with their group multiplication law.} to $SU(3)$, $\mathcal {S}=SU(3)$.

 There is also another possibility that $n=4$ but in such a case $\mathcal{S}$ is not spin.  This proves that under the global assumptions made the only horizon that preserves 6 supersymmetries has spatial horizon section diffeomeorphic to $SU(3)$.

 We have shown that  the diffeomorphic type of the spatial horizon section $\mathcal {S}$ is $SU(3)$ but we have  not specified its geometry.
It is known that $SU(3)$ admits a left-invariant, strong, HKT structure \cite{Spindel}.   A description of the strong HKT structure on $SU(3)$ as a  principal fibration over $\overline{\mathbb{C}\mathrm{P}}^2$ can be found in \cite{gp0}. So there are solutions. However, the uniqueness of the solutions remains an open problem. In particular, the question remains on whether there are spatial horizon sections preserving 6 supersymmetries,  which although  diffeomorphic to $SU(3)$, have geometry that it is not either left- or right-invariant. To answer this question will require to find the solutions of (\ref{dconx8b}) for $\mathring g$. In the $SU(3)$ case, there is a simplification. It has been shown in \cite{poon} that the only compact simply connected anti-self-dual 4-manifold with positive scalar curvature and signature one is $\overline{\mathbb{C}\mathrm{P}}^2$ and  moreover the anti-self-dual structure   has to be in the same conformal class as the Fubini-Study metric. This simplifies (\ref{dconx8b}) but still remains non-linear and it will be investigated elsewhere.
Of course if one considers $SU(3)$ with the standard left-invariant HKT structure, then more examples of spatial horizon sections can be constructed by considering the quotient $D\backslash SU(3)$, where $D$ is a discrete subgroup.

\subsubsection{Fibre group $U(1)\times SO(3)$}\label{sec:so3}

Next suppose that $\mathcal{S}$ is a principal bundle over $M^4$ with fibre group $U(1)\times SO(3)$, where $M^4$ is either  $ \#_n \overline{\mathbb{C}\mathrm{P}}^2$ or $S^4$, see (\ref{consum}).  Principal $SO(3)$ bundles over 4-dimensional manifolds are classified by their first Pontryagin class $p_1\in H^4(M^4, \bZ)$ and the second Stiefel-Whitney class\footnote{For an $SO(3)$ principal bundle to be lifted to a principal $SU(2)$ bundle, the class $w_2$ of the $SO(3)$ bundle must vanish.  This is similar to the condition required  for the existence of a spin structure on a manifold, where the $w_2$ class of the tangent bundle of the manifold must vanish.} $w_2\in H^2(M^4, \bZ)$ subject to the condition
\be
p_1= \mathcal{P} (w_2)~~\mathrm{mod}~4~,
\label{relcon}
\ee
i.e. $p_1$ and $w_2$ classes are not independent, where $\mathcal{P}$ is the Pontryagin square operation\footnote{For this view $w_2$ as class in $H^2(M^4, \bZ )$. Then $\mathcal{P}$ is the cup product of $w_2$ with itself followed by a $\mathrm{mod}~4$ operation, see explicit example below.} that takes values in $H^2(M^4, \bZ_4)$.  Thus, $\mathcal{S}$ is classified by
\be
(c_1(\mathcal{S}), w_2(\mathcal{S}), p_1(\mathcal{S}))\in H^2(M^4, \bZ)\oplus H^2(M^4,\bZ_2)\oplus H^4(M^4, \bZ)~,
\ee
subject to the relation (\ref{relcon}), where $c_1$ is the first Chern class of the principal $U(1)$ subbundle, and $w_2$ and $p_1$ are the characteritic classes of the principal $SO(3)$ subbundle.

In terms of characteristic classes, the triviality of the class $[\mathcal{F}\wedge \mathcal{F}]$ is expressed as
\be
c^2_1(\mathcal{S})+p_1(\mathcal{S})=0~.
\label{top4}
\ee
However, the fundamental representation of $SO(3)$ coincides with the adjoint representation and since the adjoint representation of $U(1)$ is trivial
\be
p_1(\mathcal{S})=p_1({\mathrm {Ad}}(\mathcal{S}))=p_1(\Lambda^+)=2\chi(M^4)+ 3\tau(M^4)~,
\ee
where again we have identified ${\mathrm {Ad}}(\mathcal{S})$ with the bundle $\Lambda^+$ of self-dual 2-forms on $M^4$.
Therefore, the condition (\ref{top4}) can be re-expressed as
\be
c^2_1(\mathcal{S})+2 \chi(M^4)+3\tau(M^4)=0~.
\label{top5con}
\ee
For $M^4= \#_n \overline{\mathbb{C}\mathrm{P}}^2$, the above condition becomes
\be
\sum_{i=1}^n m_i^2=4-n~,
\label{top5cona}
\ee
where we have used (\ref{rhsn}), expanded $c_1(\mathcal{S})=\sum_i m_i \alpha_i$  with  $\{\alpha_i; i=1,\dots, n\}$ a basis in $H^2( \#_n \overline{\mathbb{C}\mathrm{P}}^2, \bZ)$, and utilised  that the intersection matric of $\#_n \overline{\mathbb{C}\mathrm{P}}^2$ is $-{\bf 1}$.

Further progress depends on the existence of solutions to (\ref{top5cona}) by appropriately choosing $c_1$ in each case. It is clear that there are no solutions for $n=1$ while for $n=2$ and $n=3$, one has that
\begin{align}
&n=2:~~~~m_1=\pm1~,~~m_2\pm1~,
\cr
&n=3:~~~~m_1=\pm1~,~~m_2=m_3=0~,~~\mathrm{and~cyclic~in~}~~m_1, m_2, m_3~.
\end{align}
We could have also considered $n=4$ but this is ruled out because $\mathcal{S}$ is a product $\#_4 \overline{\mathbb{C}\mathrm{P}}^2\times U(1)\times SO(3)$, which is not a spin manifold, while all HKT manifolds have a spin structure.

It turns out that the principal bundle $\mathcal{S}$ with base space $\#_3 \overline{\mathbb{C}\mathrm{P}}^2$ is not a spin manifold as well.  To see this, we have to calculate the second Stiefel-Whitney class of the tangent bundle of $\mathcal{S}$, $w_2(T\mathcal{S})$. For this observe that $T\mathcal{S}$ splits into vertical $V$ and horizontal $H$  subbundles, $T\mathcal{S}=V\oplus H$, and so
\be
w_2(T\mathcal{S})=w_2(V)+w_2(H)~.
\ee
However, $V=I\oplus \mathrm{Ad}(\mathcal{S})$ and $H=\pi^* TM^4$, where $I$ is the trivial bundle and $\pi$ is the projection from $\mathcal{S}$ onto $M^4$. $\mathrm{Ad}(\mathcal{S})$ as vector bundle over $\mathcal{S}$ is trivial -- it admits a global frame basis spanned the the vector fields generated by the right $SO(3)$ free action.  Thus $w_2(V)=0$. On the other hand $H=\pi^* TM^4$.  Thus, $w_2(H)=\pi^*w_2(TM^4)$.  For $M^4=\#_3 \overline{\mathbb{C}\mathrm{P}}^2$, one has that
\be
w_2(TM^4)=\bar \alpha_1+\bar\alpha_2+\bar\alpha_3~,
\ee
where $\{\bar \alpha_i=; i=1,2,3\}$ is a basis in $H^2(\#_3 \overline{\mathbb{C}\mathrm{P}}^2, \bZ_2)=\oplus^3\bZ_2$.  To calculate the pull back of $w_2(TM^4)$ on $\mathcal{S}$, we have to specify the bundle. As $p_1(\mathcal{S})[\#_3 \overline{\mathbb{C}\mathrm{P}}^2]=1$, one has that $p_1(\mathcal{S})\,\mathrm{mod}\, 4\not=0$ and so for the relation (\ref{relcon}) to hold,  $w_2(\mathcal{S})\not=0$.  There are 8 topologically distinct $SO(3)$ principal bundles, given the first Pontryagin class, but only three satisfy the condition (\ref{relcon}).  These have  second Stiefel-Whitney class one of the basis elements $\bar \alpha_i$ of the cohomology. Choose\footnote{To calculate the Pontryagin square operation, lift the $w_2$ class to a class in $H^2(\#_3 \overline{\mathbb{C}\mathrm{P}}^2, \bZ)$ by setting $w_2=\alpha_1$. Then, $\mathcal{P}(w_2)=w_2\smile w_2= \alpha_1\smile \alpha_1=\beta\not=0$, where $\beta$ is a generator of $H^2(\#_3 \overline{\mathbb{C}\mathrm{P}}^2, \bZ)$ and $\smile$ is the cup product.} $w_2(\mathcal{S})=\bar\alpha_1$, the other two choices can be treated in a similar way. Using that the characteristic classes of principal bundles when pulled-back on the bundle space become trivial, we conclude that
\be
w_2(T\mathcal{S})=\pi^*w_2(T(\#_3 \overline{\mathbb{C}\mathrm{P}}^2))=\pi^*(\bar\alpha_1+\bar\alpha_2+\bar\alpha_3)=\pi^*(\bar\alpha_2+\bar\alpha_3)\not=0~,
\ee
and so $\mathcal{S}$ is not a spin manifold.
Thus, we have found that there are no solutions to the topological condition if the base space of the fibration is either $\overline{\mathbb{C}\mathrm{P}}^2$ or $\#_3 \overline{\mathbb{C}\mathrm{P}}^2$.

The remaining case that needs to be investigated is that of horizons with  base space $\#_2 \overline{\mathbb{C}\mathrm{P}}^2$. To see whether the associated horizons are spin manifolds, we use the analysis above described for horizons with base space $\#_3 \overline{\mathbb{C}\mathrm{P}}^2$.  Repeating the decomposition of $T\mathcal{S}$ in vertical and horizontal subspaces, $T\mathcal{S}=V\oplus H$, we again find that  $w_2(T\mathcal{S})=w_2(V)+w_2(H)=\pi^* w_2(T\#_2 \overline{\mathbb{C}\mathrm{P}}^2)$, where  $w_2(V)$ vanishes for the same reason as that given in the previous case.  It is also known that  $w_2(T\#_2 \overline{\mathbb{C}\mathrm{P}}^2)=\bar\alpha_1+\bar\alpha_2$, where $\{\bar \alpha_i; i=1,2\}$ is a basis in $H^2(\#_2 \overline{\mathbb{C}\mathrm{P}}^2, \bZ_2)=\oplus^2\bZ_2$.  It remains to see whether the pull-back of $w_2(T\#_2 \overline{\mathbb{C}\mathrm{P}}^2)$ on $\mathcal{S}$ vanishes.
For this observe that $p_1(\mathcal{S})[\#_2 \overline{\mathbb{C}\mathrm{P}}^2]=2\, \mathrm{mod}\, 4=2\not=0$.  Thus, for the relation (\ref{relcon}) to hold,  $w_2(\mathcal{S})\not=0$.  Moreover, the Pontryagin square of $w_2$ when evaluated on   $\#_2 \overline{\mathbb{C}\mathrm{P}}^2$  must also give $ 2\, \mathrm{mod} \, 4$. The only possibility is that $w_2(\mathcal{S})=\bar\alpha_1+\bar\alpha_2=w_2(T\#_2 \overline{\mathbb{C}\mathrm{P}}^2)$.  As the characteristic classes of principal bundles when they are pulled back on the bundle space become trivial, we conclude that $w_2(T\mathcal{S})=0$ and so $\mathcal{S}$ is spin. This solves the topological condition required  for the existence of a horizon geometry.

Despite the complexity of the bundle construction of $\mathcal{S}$ described above, $\mathcal{S}$ is a product of ``simple'' manifolds. In particular, it is a consequence of the Smale-Barden classification of 5-dimensional compact simply connected manifolds that $\mathcal {S}$ is diffeomorphic to $S^2\times S^3\times SO(3)$.  To see this, view $\mathcal{S}$ as a two stage fibration, first as a circle bundle $N^5$ over $\#_2 \overline{\mathbb{C}\mathrm{P}}^2$ with $c_1=\alpha_1+\alpha_2$ and then a bundle over $N^5$ with fibre $SO(3)$. It turns out that $N^5$ is simply connected. One can compute the second homology class of $N^5$ to find $H_2(N^5, \bZ)=\bZ$. This implies that $N^5$ is diffeomorphic to $S^2\times S^3$. As $H^4(S^2\times S^3, \bZ)=0$, the pull back of $p_1(\mathcal {S})$ on $N^5$ vanishes. Then, the relation (\ref{relcon}) also implies that the pull back of $w_2$ on $N^5$ vanishes as well.  Thus $\mathcal{S}$ is a topologically trivial bundle over $N^5$, which proves the statement.

The manifold $\#_2 \overline{\mathbb{C}\mathrm{P}}$ admits several anti-self-dual structures constructed in \cite{poon, donaldsonfriedman, lebrun2}. Some of these have positive scalar curvature and so they can be candidates for inducing the required geometric structure on $S^2\times S^3\times SO(3)$ to lead to horizon solution that preserves 6 supersymmetries. However, it still remains to identify whether one of these anti-self-dual structures on  $\#_2 \overline{\mathbb{C}\mathrm{P}}$
solves (\ref{dconx8b}). Incidentally, this will also give an example of a compact HKT manifold with torsion a closed 3-form.  This question will be  examined elsewhere.

\section{Heterotic AdS$_3$ backgrounds}\label{Perelmanstyle}

\subsection{Geometry of supersymmetric AdS$_3$ backgrounds}

It has been demonstrated in \cite{sbjggp} that supersymmetric AdS$_3$ backgrounds are (direct) products, AdS$_3\times M^7$, i.e. the warped factor is constant, where the 7-dimensional manifold $M^7$ admits at most an $SU(3)$ structure. In particular, the spacetime metric $g$ and 3-from field strength $H$ are given by
\be
g= 2 e^+ e^-+  dz^2+ \bar g~,~~~H= X e^+\wedge e^-\wedge dz+ \bar H~,~~~
\label{ads3gh}
\ee
respectively, where
\be
e^+=du~,~~~e^-=dr-\frac{2r}{\ell} dz~,
\ee
  the wrapped factor $A$ in \cite{sbjggp}, without loss of generality,  has been set to one, $A=1$ and $X$ is constant.  It can be seen after a coordinate transformation that $g_{\mathrm{AdS}}\equiv 2 e^+ e^-+ dz^2$ is the metric on AdS$_3$ of radius $\ell$.  Moreover, $\bar g$ and $\bar H$, with $\bar H$ closed, $d\bar H=0$,  are the metric and 3-form torsion\footnote{The notation $\bar g$ and $\bar H$ does not denote complex conjugation of $g$ and $H$ as all these tensors are real. Instead, it is used to distinguish the  metric and 3-form field strength of the transverse space $M^7$ from those of the spacetime.}  on the transverse space $M^7$, respectively,  and they are independent from the AdS$_3$ coordinates $(u, r, z)$.  Similarly, the dilaton $\Phi$ is a function of $M^7$.

The spacetime metric $g$ and 3-form field strength $H$ in (\ref{ads3gh}) are a special case of those of the black hole horizon background in (\ref{horgh}) provided that the spatial horizon section is taken to be non-compact, $\mathcal S= \bR\times M^7$ and
\be
h=-\frac{2}{\ell} dz~.
\label{adsh}
\ee
Therefore, $dh=0$ and $k^2=4 \ell^{-2}$.

In what follows, we shall also use the expression for the Ricci scalar of $M^7$
\be
R(\bar g)=\frac{1}{4} \bar H^2-2 \bar\nabla^2 \Phi~,
\label{eineqn}
\ee
where $\bar\nabla$ is the Levi-Civita connection of $\bar g$.
This follows  upon substituting (\ref{ads3gh}) into  the Einstein equation of the heterotic theory.  Moreover, the closure of $\bar H$, $d\bar H=0$, is implied by that of $H$.

\subsection{AdS$_3$ backgrounds with 6 supersymmetries  }\label{preliminaries}

\subsubsection{Summary of geometric and global conditions}

It has been demonstrated in \cite{sbjggp} that for AdS$_3$ backgrounds that preserve 6 supersymmetries, $M^7$ admits three $\h{\bar \nabla}$-covariantly constant vector fields, whose Lie algebra is $\mathfrak{su}(2)$,  where $\h{\bar \nabla}$ is the metric connection on $M^7$ with torsion $\bar H$. Denoting, after an orthonormal normalisation, the associated dual 1-forms with $\lambda^r$, $r=1,2,3$, one finds that
\be
\bar g=\delta_{rs} \lambda^r \lambda^s+ e^{2\Phi} \mathring g~,~~~\bar H=\mathrm{CS} (\lambda)-\mathring{*} de^{2\Phi}~,
\label{fghads}
\ee
where $\mathring g$ is the metric on the space of orbits $M^4$ of the vector fields and $\Phi$ is the dilaton that depends only on the coordinates of $M^4$. From now on, viewing $M^{7}$ as a principal bundle over $M^4$ with fibre either $SU(2)$ or $SO(3)$, $\lambda^r$ can be interpreted as a principal bundle connection and $\mathrm{CS}(\lambda)$ in (\ref{fghads})  is the Chern-Simons 3-form of $\lambda$.
Moreover, $(M^4, \mathring g, \mathring{I}_r)$ is a quaternionic K\"ahler manifold, i.e. it satisfies
\be
\mathring \nabla \mathring{I}_r+\frac{k}{2} \xi^t \epsilon_t{}^s{}_r \mathring{I}_s=0~,
\label{qkxx}
\ee
 whose geometry  will be restricted further later, where $\xi$ is the pull-back of $\lambda$ on $M^{4}$ with a local section  and $k$ is given in terms of the radius of AdS$_3$ below eqn (\ref{adsh}).
  The KSEs restrict  the self-dual part of the curvature $\mathcal{F}$ of $\lambda$ to satisfy
\be
(\mathcal{F}^r)^\mathrm{sd}=\frac{k}{4} e^{2\Phi} \mathring\omega^r~,
\label{sdf}
\ee
where $\mathring \omega^r$ is the Hermitian form associated to the $\mathring{I}_r$ and $\mathring g$.
The closure of $\bar H$ implies two conditions. One is the topological condition that the cohomology class
\be
[\delta_{rs} \mathcal{F}^r\wedge \mathcal{F}^s]=0~,
\label{topcon}
\ee
must be trivial and the other is the differential condition that
\be
\mathring{\nabla}^2 e^{2\Phi}=\frac{3 k^2}{8} e^{4\Phi}-\frac{1}{2} \sum_r\Big((\mathcal{F}^r)^{\mathrm{asd}}\big)_o^2~,
\label{dileqn}
\ee
where $\mathcal{F}^{\mathrm{asd}}$ is the anti-self-dual component of the curvature of $\lambda$.

\subsubsection{Non-existence of smooth solutions}

The proof of this non-existence  result is similar to the existence proof described in section \ref{sec:hor} for horizons.  The difference is that $M^7$ is an $SU(2)$ or $SO(3)$ principal bundle over $M^4$ and so there is not an analogue of the line bundle $L$, which is necessary to find a solution to the topological condition (\ref{topcon}).  So, we shall describe this proof very briefly.

First a comparison of (\ref{eineqn}),  which arises from the Einstein equation, with that in (\ref{dileqn}), which arises from the closure of $\bar H$, yields
\be
 \mathring R=\frac{3 k^2}{2} e^{2\Phi}>0~,
 \label{pR}
 \ee
i.e. the scalar curvature of $(M^4, \mathring g, I_r)$ is positive. This is analogous to the equation (\ref{posR}) for horizons. In addition, using this equation and an argument similar to that presented for horizons, one can  recast (\ref{dileqn}) in terms of the Ricci tensor and scalar as in (\ref{dconx8b}). Moreover, a  consequence of (\ref{qkxx}), (\ref{sdf}) and (\ref{pR}) is that $M^{4}$ is an anti-self-dual 4-manifold with positive scalar curvature, see again proposition 7.1, page 92 in \cite{salamon}. Thus as for horizons, it is a consequence of the results in  \cite{lebrun} that $M^4$ is homeomorphic to either $S^4$ or the connected sum $\#_n{\overline{\mathbb{C}\mathrm{P}}}^2$, see (\ref{consum}).

Viewing $M^7$ as a principal bundle with fibre group $SU(2)$, it is a consequence of (\ref{qkxx}) that
its associated adjoint bundle $\mathrm{Ad}(M^7)$ should be identified with the vector bundle of self-dual 2-forms on $M^4$, i.e.
\be
\mathrm{Ad}(M^7)=\Lambda^+(M^4)~.
  \ee
  Such principal bundles are classified by the second Chern class of the associated fundamental vector bundle $E$, $c_2(E)\in H^4(M^4, \bZ)$ as $c_1(E)=0$.  This allows us to the express the first Pontryagin class of $\mathrm{Ad}(M^7)$ as
\be
p_1(\mathrm{Ad}(M^7))=p_1(\Lambda^+)=2\chi+3\tau=-4 c_2(E)~,
\label{pone}
\ee
where the second equality follows from the Hirzenbruch's signature theorem, as for horizons, and the third equality from the classic formula (\ref{classf}) with $c_1(E)=0$ -- $\chi$ is the Euler characteristic class and $\tau$ is that of the signature class of $M^4$.

 After an appropriate overall normalisation, the class
\be
[\delta_{rs} \mathcal{F}^r\wedge \mathcal{F}^s]=p_1(E)=-2 c_2(E)=\frac{1}{2} \Big(2\chi+3\tau\Big)~,
\ee
where we have used (\ref{pone}),  the relation,  $p_1(E)=c_1(E)^2-2 c_2(E)=-2c_2(E)$, between the characteristic classes of $E$ and $c_1(E)=0$.  As a result, the topological condition (\ref{topcon}) implies that
\be
(2\chi+3\tau)[M^4]=0~.
\label{top4}
\ee
Clearly, for $M^4= S^4$ the above topological condition cannot be satisfied as the Euler number of this space  is 2 and the signature vanishes. For the rest of the possibilities $(2\chi+3\tau)[\#_n {\bar{\mathbb{C}\mathrm{P}}}^2]=4-n$.  The topological condition is satisfied for $n=4$.  But this is not an acceptable solution because $M^7= \#_4 {\bar{\mathbb{C}\mathrm{P}}}^2\times S^3$ and such a space does not admit a spin structure.  

A similar analysis can be done if $M^7$ is considered as a principal $SO(3)$ bundle over a simply connected 4-dimensional manifold $M^4$. Again, $M^4$ must be homeomorphic to either $S^4$ or $\#_n {\bar{\mathbb{C}\mathrm{P}}}^2$.  The topological condition implies that $2\chi+3\tau$ must be a trivial class -- the argument is similar to that presented in section \ref{sec:so3}, see eqn (\ref{top5con}) with $c_1=0$.   Therefore, solutions cannot exist as a consequence of the same argument as that presented for the $SU(2)$ case above. Thus, there do not exist smooth AdS$_3$ backgrounds
with compact transverse space for which the $\mathfrak{su}(2)$ action can be integrated to a free group action with a simply connected orbit space that admit 6 supersymmetries.

\section{Horizons and AdS solutions with 4 supersymmetries revisited} \label{sec:four}

\subsection{Geometry of horizons and AdS solutions}

Having established our results for horizons and AdS$_3$ backgrounds that preserve 6 supersymmetries, we shall comment on some of the properties of such backgrounds that preserve 4 supersymmetries. The geometric conditions required for the existence of such backgrounds have been presented\footnote{The notation that we use here has some differences from that in \cite{jggphh}) and \cite{jggp2}. But the differences are self-explanatory.} in \cite{jggphh} and further explored in \cite{jggp2}.  Here, we shall not describe the general case. Instead, we shall focus on the horizons for which the spatial section $\mathcal {S}$ is a $T^4$ bundle over a 4-dimensional manifold $M^4$.  The metric and 3-form field strength of $\mathcal{S}$ can be written as
\be
\tilde g=\frac{1}{k^2} \Big( \sum_{r=1}^3 (h^r)^2+ w^2\Big) + e^{2\Phi}\mathring g~,~~~\tilde H=\mathrm{CS}(\lambda)-\mathring * de^{2\Phi}~,
\ee
where $(\lambda^0, \lambda^1, \lambda^2, \lambda^3)=k^{-1}( w, h^1,  h^2,  h^3)$ are the components of a principal bundle connection  with $h^1=h$ and  $w^2= (h^1)^2= (h^2)^2=(h^3)^2=k^2$ have constant length $k^2$. Moreover, the conditions that the geometry satisfies, put in form notation,  can be expressed as
\bea
&&dh^r\wedge \mathring {\omega}=0~,~~~r=1,2,3~;~~~dw\wedge \mathring {\omega}=-{k^2\over2}\, e^{2\Phi} \mathring {\omega}\wedge \mathring\omega~,~~~
\cr
&&\rho(\mathring \omega)\equiv -i\partial\bar\partial \log \det(i\mathring {\omega})= dw~, ~~~ dw^{2,0}=dw^{0,2}=(dh^r)^{2,0}=(dh^r)^{0,2}=0~,
\cr
&&d\mathrm{CS}(\lambda)+2 i
\partial\bar\partial e^{2\Phi}\wedge \mathring {\omega}\equiv k^{-2}\sum_{r=1}^3 dh^r\wedge dh^r +k^{-2} dw\wedge dw+2 i
\partial\bar\partial e^{2\Phi}\wedge \mathring {\omega}=0~,
\label{allconk}
\eea
see equation (4.2) in \cite{jggp2}, where $M^4$ is a K\"ahler manifold with metric $\mathring g$ and K\"ahler form $\mathring\omega$, $d\mathring {\omega}=0$, $\rho(\mathring \omega)$ is the Ricci form, and $\partial$ and $\bar\partial$ are the holomorphic and anti-holomorphic exterior derivatives on $M^4$, respectively. Note that $dh^r$ are anti-self-dual 2-forms while $dw$ satisfies a Hermitian-Einstein type of condition, i.e. all of them are $(1,1)$-forms on $M^4$.

It is remarkable that a differential system very similar to (\ref{allconk}) has recently arisen in the classification of 6-dimensional compact strong Calabi-Yau manifolds with torsion in \cite{abls}. In particular, the differential system of \cite{abls} can be written as in (\ref{allconk}) after setting $h=h^1$ and $h^2=h^3=0$. In such a case, $\mathcal{S}$ becomes a compact strong 6-dimensional Calabi-Yau manifold with torsion. The  vector fields  associated to $h$ and $w$ do not emerge  as a consequence of the  KSEs, see  \cite{jggphh}, but instead they arise from the fact that strong compact Calabi-Yau manifolds with torsion are gradient generalised Ricci solitions, see e.g. \cite{gpew, abls} and references therein. If these spaces have non-vanishing torsion, they admit two $\hat{\tilde \nabla}$-covariantly constant vector fields.  In turn, these vector fields give rise to  $h$ and  $w$.

The third condition in (\ref{allconk}) implies that the Ricci-form of K\"ahler geometry on $M^4$ is equal to the curvature of a line bundle, which is a (1,1)-form.  Therefore this bundle is the canonical bundle of $M^4$. Using the second condition on $dw$ as well as the third condition, we conclude that
\be
\mathring R=\frac{ k^2}{2} e^{2\Phi}>0~.
\label{rphi2}
\ee
Thus $M^4$ is a K\"ahler manifold with positive scalar curvature.

The first condition in (\ref{allconk}) can always be solved for each $r$. Indeed,
consider the fundamental complex line bundle $L$ associated to a circle principal bundle $P$ on a K\"ahler manifold $M^4$ and let us assume that it is
holomorphic -- this is equivalent to requiring that it admits a connection whose curvature is a (1,1)-form with respect to the complex structure on $M^4$. Then, it can be shown that one can always find another connection such that its curvature satisfies $F\wedge \mathring \omega=0$ provided that the degree of the line bundle vanishes, i.e.
\be
\mathrm{deg}(L)\equiv (c_1(L)\wedge \mathring \omega)[M^4]=0~.
\label{antopcon}
\ee
 To prove this, suppose that $F=dA$ and consider another connection $A'=A+ \iota_I df= A+ i(\partial-\bar\partial)f$, where $f$ is a function on $M^4$. Imposing the condition $F(A')\wedge \mathring \omega=0$ on $A'$,  one finds that
 \be
 \mathring\nabla^2 f+\frac{1}{2} \mathring \omega \cdot F(A)=0~,
 \ee
 where $\mathring\omega \cdot F(A)$ denotes the $\mathring\omega$-trace of $F(A)$.
 This equation can be inverted for $f$ provided that $\omega \cdot F(A)$ does not  have a harmonic component. The latter is equivalent to the condition (\ref{antopcon}). Therefore, the first condition in (\ref{allconk}) can be solved provided that the line bundles with curvature $\mathcal {F}^r=k^{-1} d h^r$ are holomorphic and have zero degree -- the curvatures $\mathcal {F}^r$ are anti-self-dual 2-forms on $M^4$.  The associated cohomology classes in $H^2(M^4, \bZ)$  are often referred to as {\it primitive}.

 Moreover, the third condition in (\ref{allconk}) can be solved as well.  It is sufficient for this condition to be satisfied provided  that one of  the fundamental complex line bundles of  the principal $T^4$ fibration  is identified with the canonical bundle of $M^4$. The necessary condition can be weaker as it only required that the canonical bundle of $M^4$ is one of the associated complex line bundles of the $T^4$ principal fibration -- we shall illustrate this with an example below.  As the canonical bundle of a K\"ahler manifold is holomorphic, e.g. with respect to the connection induced from the K\"ahler metric of $M^4$, there is always a connection $w$ that satisfies the third condition in (\ref{allconk}).

 The last condition in (\ref{allconk}), which is the closure of $\tilde H$,  gives raise to a topological condition and a differential one on $M^4$. The former has also been investigated in \cite{jggp2}. If $\mathcal {F}^0\equiv k^{-1} dw$ has been identified with the curvature of the canonical bundle, then Wu's formula implies that
 \be
 c_1^2=2\chi+3\tau~,
 \ee
 where $c_1$ is the first Chern class of $M^4$, $\chi$ is the Euler class and $\tau$ is the signature class of $M^4$.  Given a basis $\{E_\alpha ; \alpha=1, \dots, m\}$ of primitive cohomology classes in $H^{1,1}(M^4, \bZ)$, i.e cohomology classes associated with holomorphic complex line bundles of zero degree, the cohomology condition that arises from the closure of $\tilde H$ can be put into the form
 \be
 \sum_r n_r^\alpha n_r^\beta E_\alpha\cdot E_\beta+2\chi[M^4]+3\tau[M^4]=0~,
 \label{top5}
 \ee
 where $(E_\alpha\cdot E_\beta)$ is the intersection matrix of the basis and $\{n_r^\alpha\in \bZ; \alpha=1,\dots, m; r=1,2,3\}$, see \cite{jggp2}. Given $M^4$ and therefore the Euler number and the signature of the space, (\ref{top5}) is considered as an equation for the integers $n_r^\alpha$. For the topological condition to hold, there must be such solutions.

 The differential condition that arises from the last condition in (\ref{allconk}) can be cast into the form of (\ref{pde}) as
\be
\mathring{\nabla}^2 e^{2\Phi}
=\frac{k^2 }{8}  e^{4\Phi} -\frac{1}{2 k^2} \Big(\sum_r(dh^{r})_o^2+ (dw^{\mathrm{asd}})_o^2\Big)~,
\label{dconxx}
\ee
 where $u=e^{2\Phi}$. It has been demonstrated in \cite{gp0} that this equation can always be solved for $e^{2\Phi}$.  However, this is not the equation that we should be solving. The equation that has to be solved is derived from (\ref{dconxx}) upon substituting (\ref{rphi2}) into (\ref{dconxx}) and after expressing $dw^{\mathrm{asd}}$ in terms of the Ricci tensor of the underlying K\"ahler geometry using the third equation in (\ref{allconk}) -- the components of Ricci form $\rho$ in complex coordinates  are given in terms of those of the Ricci tensor. The final expression reads
\be
\mathring{\nabla}^2 \mathring R
=\frac{5 }{16}  \mathring R^2-\frac{1}{4} \mathring R_{ij}\mathring R^{ij}-\frac{1  }{4} \sum_r(dh^{r})_o^2 ~,
\label{dcon3x}
\ee
where the indices have been raised with $\mathring g$.
 This equation should be thought as the equation that determines the K\"ahler metric $\mathring g$ on $M^4$.  Solutions to this equation have been described in \cite{jggp2}. But it is not known whether it always admits a solution on a K\"ahler manifold $M^4$. For example, one question is whether given a K\"ahler metric $\mathring g$ on $M^4$, one can find another one $\mathring g'$, such that $\mathring \omega$  and $\mathring \omega'$ have the same cohomological class,  that solves (\ref{dcon3x}). As they are in the same cohomological class, it is a consequence of the $\partial\bar\partial$-lemma that $\mathring \omega'=\mathring\omega+i\partial\bar\partial f$ for some function $f$ on $M^4$. Then, (\ref{dcon3x})  becomes a non-linear PDE on $f$ and it is not known whether it can be solved in general.

All solutions $\mathcal{S}$ to (\ref{allconk}), including those of the topological condition (\ref{top5}), do not exhaust all possible solutions to the system. This is the case whenever $\mathcal{S}$ is not simply connected. In such a case,  one has to also consider all appropriate covers of the solutions obtained. This is because all such covers exhibit the same local geometry as the original solutions and so they are solutions themselves. This also applies to the AdS$_3$ backgrounds and it will be illustrated with an example below.

The above analysis can be repeated for AdS$_3$ backgrounds that preserve four supersymmetries. For this suffices to set
\be
h=h^1=-\frac{2}{\ell} dz~,
\ee
where $\ell$ is the radius of AdS$_3$.
As $dh=0$, its contribution vanishes in all the formulae in (\ref{allconk}). Otherwise, the analysis can be carried out as above. Again the conclusion is that the existence of solutions requires for
(\ref{dcon3x})  to admit solutions  for some K\"ahler metric on $M^4$, where now only $dh^2$ and $dh^3$ contribute in the sum over $r$ as $dh^1=0$.

\subsection{An example}

It is instructive to pursue an example. For this, let us consider the well-known AdS$_3$ solution AdS$_3\times S^3\times S^3\times S^1$ widely used in AdS/CFT \cite{boonstra}-\cite{witten}. The transverse space is $M^7=S^3\times S^3\times S^1$. Clearly $M^7$ is a principal $T^3$ fibration over $M^4=S^2\times S^2$, which is, as expected,  a 4-dimensional K\"ahler manifold.

 To reverse engineer the construction starting from the base space $M^4=S^2\times S^2$,   the Euler number of $S^2\times S^2$ is $\chi[S^2\times S^2]=4$ and the signature vanishes $\tau[S^2\times S^2]=0$. The generators of $H^2(S^2\times S^2, \bZ)=\bZ\oplus \bZ$ can be represented by forms  $\alpha$ and $\beta$ such that the intersection matrix is $\alpha\cdot \beta=\beta\cdot \alpha=1$ with $\alpha^2=\beta^2=0$, e.g. $\alpha$ and $\beta$ can be the normalised volume forms of the 2-spheres.

 The first Chern class $c_1$ of the canonical bundle of $S^2\times S^2$ can be represented by $c_1=2\alpha+2\beta$.  Indeed, using the intersection matrix
 \be
 c_1^2[S^2\times S^2]= 8= 2 \chi[S^2\times S^2]+3\tau[S^2\times S^2]~,
 \ee
 which is Wu's formula.

The metric on $S^2\times S^2$ can be taken as the sum of two Fubini-Study metrics (up to an overall scale) one for each $S^2$ subspace. The K\"ahler form $\mathring \omega $ can be chosen such that
\be
\mathring \omega=\alpha+\beta~.
\ee
This is a convenient choice for the total volume of the space to be $1$ as $d\mathrm{vol}=1/2 \mathring\omega\wedge \mathring\omega$.  But one can also consider any multiple  $r(\alpha+\beta)$, $r\in\bR_{>0}$. Furthermore to solve the topological condition (\ref{top5}), one can choose
\be
dh^2= 2 \alpha-2\beta~,~~~ dh^3=0~.
\ee
With these choices, the topological condition (\ref{top5}) is satisfied as well as all the rest of the conditions in (\ref{allconk}). Clearly, the transverse space of this AdS$_3$ solution is a product,  $M^7=Q\times S^1$,  as one of the first Chern classes of the fibration $T^3\hookrightarrow M^7\rightarrow S^2\times S^2$ vanishes ($dh^3=0$).  However, the fibration
\be
T^2\hookrightarrow Q\rightarrow S^2\times S^2~,
\ee
with first Chern classes $2\alpha+2\beta$ and $2\alpha-2\beta$ has bundle space $Q$
\be
Q=S^3\times S^3/\bZ_4\oplus\bZ_2~,
\ee
and not $S^3\times S^3$.  To outline  a proof for this, the Chern classes lifted to the associated principal bundle become trivial. Thus here, they give the relations $2\alpha+2\beta=0$ and  $2\alpha-2\beta=0$ on $Q$. These can be solved by $\beta=x$ and $\alpha=x+y$, where $x$ is the generator of $\bZ_4$ and $y$ is the generator of $\bZ_2$.  As a result, the generators $x,y$ ``survive'' when lifted to the bundle space and generate $\bZ_4\oplus\bZ_2$, which becomes the fundamental group of $Q$. (There is an either spectral sequences  for fibrations argument or an argument based on exact homotopy sequences for fibrations to establish this.)

Clearly, $S^3\times S^3$ is the universal cover of $Q$. It is known that given a manifold $M$ and a discrete group $D$, the geometry of $M/D$, like metric, forms and complex structure, can be lifted to $M$, especially if $D$ is a finite group as in the case at hand. Thus, the solution AdS$_3\times S^3\times S^3\times S^1$ can be recovered from that of AdS$_3\times Q\times S^1$ upon considering the universal cover of $Q$.  Therefore, to find all possible such heterotic backgrounds,  it is necessary to also consider the covers of the solutions obtained by solving the conditions (\ref{allconk}).



 \bibliographystyle{unsrt}

\end{document}